\documentclass[aps,twocolumn,showpacs]{revtex4}

\begin{document}

\title{Unified Theory of Interspecific Allometric Scaling}

\author{Jafferson K. L. da Silva, Lauro A. Barbosa$^*$ and Paulo Roberto Silva}

\affiliation{
Departamento de F\'\i sica,Instituto de Ci\^encias Exatas,
 Universidade Federal de Minas Gerais\\
C. P. 702, 30123-970, Belo Horizonte, MG, Brazil\\
$^*$Present address: Instituto de F\'\i sica de S\~ao Carlos, Universidade de S\~ao Paulo,\\
        Caixa Postal 369, 13560-970 S\~ao Carlos, S\~ao Paulo, Brazil\\
        E-mail: jaff@fisica.ufmg.br, lau@ifsc.usp.br and prsilva@fisica.ufmg.br}

\date{\today}\widetext

\pacs{87.10.+e, 87.23.-n}

\begin{abstract}

 A general simple theory for the interspecific allometric scaling is developed
 in the $d+1$-dimensional space ($d$ biological lengths and a physiological time) of
 metabolic states of organisms. It is assumed that natural selection shaped the metabolic states 
 in such a way that the mass and energy $d+1$-densities are
 size-invariant quantities (independent of body mass). 
 The different metabolic states (basal and maximum) are described
  by considering that the biological lengths and the physiological time are related by 
  different transport processes of energy and mass. 
  In the basal metabolism, transportation occurs by ballistic and diffusion processes.
  In $d=3$, the $3/4$ law occurs if the ballistic movement is the dominant process, while
  the $2/3$ law appears when both transport processes are equivalent. 
  Accelerated movement during the biological
  time is related to the maximum aerobic sustained metabolism, which is characterized 
   by the scaling exponent $2d/(2d+1)$ ($6/7$ in $d=3$). 
   The results are in good agreement with empirical data and a verifiable empirical prediction
   about the aorta blood velocity in maximum metabolic rate conditions is made.

\end{abstract}
 
\maketitle

 Metabolic rate $B$ and body mass $M$ are connected by the
 relation $B=aM^b$, where $b$ is the allometric exponent and $a$ is a constant. 
 For several decades it was accepted that the basal metabolic rate  (BMR)
 among almost all organisms \cite{1,2,3} (interspecific scaling) was characterized by $b=3/4$ 
 (Kleiber's law \cite{4}). A few years ago, theoretical explanations of the 
 ubiquity of the $3/4$-law based on the resource distribution network common 
 to all organisms \cite{5} and on network geometry optimization \cite{6,banavar2002} 
 were proposed. Kleiber's law has however been questioned recently.
 On the observational side, it is not clear whether the value of $b$ 
 is $3/4$, $2/3$ or even variable in both interspecific and
 intraspecific (same species) scaling \cite{7,8,9}. 
 On the theoretical side, there are several debates about the validity of these 
models \cite{7,9b} undermining the rational basis for
 the scaling law.
  A related open question is why maximum aerobic sustained metabolic rate (MMR) of endothermic 
  animals scales with an exponent larger than that of BMR \cite{8,9,9b,10,11,12}.
  In this work we show how, on the basis of a few quite general hypotheses, all
 the aspects of metabolic scaling mentioned above may be accounted for.
  The central assumption of our approach is that, if one characterizes the metabolic
 state of an organism as a point in a space comprised of $d$ biological lengths
 and a biological time $\tau$, then  the $(d+1)$-dimensional mass and energy
 densities are size-invariant under natural selection.  Using this idea
 we may, given the dominant mass and energy transport processes, obtain
 dynamical relations involving lengths and time. The latter are
 characterized by invariant quantities (diffusion coefficient, velocity etc.). 
 Each transport mechanism is related to a different metabolic
 state, with its own values for allometric exponents, in agreement with data.

 The key questions in interspecific allometry are:
   (1) Is there a universal interspecific BMR exponent $b$?
   (2) Is there a model that can describe both BMR and MMR scaling?
   (3) Which quantities have their scaling determined by
       the BMR, and which by the MMR?
   
 We present a simple and general theory that  answers, at least  partially, these questions 
 and explain some aspects of the allometric scaling of organisms. 
 As all biological processes depend on characteristic times, it is natural to include
 a characteristic time, as well as various 
 characteristic lengths, when specifying the metabolic state of an organism \cite{13,14}.
 We therefore associate the metabolic state of an organism with
  a point in a $d+1$-dimensional space with $d$ biological lengths $L_1,~L_2\ldots,~L_d$ 
 and a biological time $\tau$. Here $d$ is the number of spatial dimensions;
  although we usually have $d=3$, we will work in  general dimension $d$. 
  Examples of biological lengths $L$ and times $\tau$  are the total aorta length in mammals,
  the length of capillaries, 
  the mean distance from cell surface to mitochondria in unicellular organisms, the 
  duration of one heartbeat, the capillary blood transit time or
  the turnover time for glucose \cite{1,2,banavar2002,14}. 

 We now look for simple, general relations constraining the distribution of
  points in the space of metabolic states. It is usual for
 an animal to make several transitions between states $A_{basal}$ 
 and  $A_{max}$, through a series of complex biochemical processes. We do not consider
 these transitions. Rather we are interested in describing how 
 natural selection has shaped the state $A_{basal}$ and other ones belonging to the 
 space of metabolic states of  all organisms (or a group of them).
  An organism is characterized by the fundamental quantities of mass and  available energy for
  the metabolic processes.
  We therefore identify the mass $M$ and available energy $E$ as the fundamental variables 
 characterizing an organism.
   Since we have a $d+1$-dimensional space we  shall use the mass density  
  $\rho_{d+1}(L_1,\ldots,L_d,\tau)$ (mass per unit volume and unit time)
 and the  energy density $\sigma_{d+1}(L_1,\ldots,L_d,\tau)$ (available energy
  per unit volume and unit time).
  We assume that during evolution,  natural selection enforces the constraints of {\sl
 size-invariant} (independent of body mass) $\rho_{d+1}$  and $\sigma_{d+1}$ (our first and second
 hypotheses, respectively). Our third hypothesis is that the scaling
 of the metabolic states is determined by the {\sl dominant dynamical transport
 processes of nutrients} (mass and energy), which are characterized by  size-invariant 
 quantities (diffusion coefficient, average velocity, etc.). 
  Note that the first and second hypotheses  
 furnish two relations  valid for all  metabolic regimes.  
 Different metabolic scalings will appear  because there are different ways to transport 
 nutrients. 
 
 From the second hypothesis  we obtain that $E=\sigma_{d+1}~\tau V_d$, 
 where $\tau V_d$ is the $d+1$-volume 
 and $V_d=L_1L_2\ldots L_d$.
 Since power is defined as $P=dE/dt$, energy can be written in terms of $B$,
 the power averaged over the time scale $\tau$, as $E=B\tau$. We identify $B$ as the
 metabolic rate.  Therefore from the first and second hypotheses  we obtain that
\begin{equation}
  M=\rho_{d+1}~\tau V_d,~~~{\rm and}~~~B=\sigma_{d+1}~ V_d~~.\label{eq1}
\end{equation}
 Let us briefly present a qualitative argument about an optimal delivery that supports our postulates.
 Consider two organisms with the same body mass $M$ belonging to the same group. 
 The organism with the larger biological volume $V_d$ has 
 the larger nutrient distribution network and more fuel and oxygen are arriving to 
the cells in a unit time. Therefore its cells must have a fast metabolism in 
order to consume the fuel. Of course, a fast metabolism is related to a small 
biological time. On the other hand, a small $V_d$ implies in a larger biological 
time $\tau$. These arguments suggest that the product $\tau V_d$ is constant for 
these organisms.

 An immediate consequence of our hypotheses is that the power
 per mass (specific metabolic rate), namely 
\[
\frac{B}{M}=\frac{\sigma_{d+1}}{\rho_{d+1}}\frac{1}{\tau}~~,
\]
 is inversely proportional to the
 metabolic time. Animals with a small $\tau$, such as small mammals,
 require  larger power by unit mass than ones with a large $\tau$ (large mammals) because
 their cells have a large mitochondrial density \cite{19}.

 Obviously, $\tau$ cannot be $0$ neither $\infty$; there must be a minimal  and a maximal
 metabolic times $\tau_{min}$ and $\tau_{max}$.
 For animals, a lower bound for $\tau_{min}$ can be found from the observation that 
 the biological volume of an organism $V_d$ cannot be larger than its {\sl spatial volume} $V$, 
 since the biological
 lengths $L_i$ characterize the organism's anatomy on the scale of the body, or
 some organ or cellular structure.
  For compact animal bodies we have that $M=\rho V$, where
 $\rho$ is the usual $d$-dimensional mass density, which is approximately constant. 
 Using $\rho V=M=\rho_{d+1}~\tau V_d$ in  $V>V_d$, 
 we obtain  $\tau>\tau_{min}=\rho/\rho_{d+1}$. 
 Since $B/M$ cannot be zero it must exist an size-invariant $(B/M)_{min}$ related
 to the minimum power per mass to keep the organisms just alive, the so-called tissue maintenance
 specific metabolic rate \cite{9}. This implies that
 $\tau_{max}= \sigma_{d+1}/(\rho_{d+1}(B/M)_{min})$.
  Note if $\tau$ is size-invariant, such as $\tau_{max}$, we have isometric 
scaling of the metabolic rate ($b=1$).

 The relation $M\propto \tau V_d$, derived from the first hypothesis,
 is a generalization of the result of Banavar et al. \cite{6}, namely $V_{net}\propto L^{d+1}$, where 
 $V_{net}$ is the total volume of an  efficient distributive network. Using that the blood
 volume $V_{net}$ is proportional to mass, they obtained $M\propto L^{d+1}$, a basic relation
 to deduce the $3/4$-law. The two relations are equal when $\tau\propto L$, a condition valid
  for the BMR. Moreover, the result $M\propto L^{d+1}$ is
  also crucial to obtain the BMR exponent in the model of West, Brown and Enquist (WBE) \cite{5}. 
This relation is also a generalization of the equation $F \propto (L_p/u)B$ 
of Banavar et al. \cite{banavar2002}, where $L_p$ is the physical length of the system 
and $u$ is the characteristic length scale. If we rewrite this equation as 
$M \propto \rho t V_d$, where $\rho$ is the tissue density and $t$ is the physiological time 
related with the rate of energy use per unit volume, it becomes similar to Eq. (\ref{eq1}).

  Although, from our third assumption, we need some dynamical size-invariant quantities, 
  like the blood flow speed velocity $v_0$ in the aorta or in the capillaries,
  the length $l_c$ and the radius $r_c$ of capillaries 
  {\sl are not necessarily invariants}.

 Let us first study the case of  transport via diffusion.
  We have only one metabolic length scale 
 ($L_1\propto L_2\propto\ldots\propto L_d\propto L$), so
  the biological volume is given by $V_d\propto L^d$.
  Since diffusion over short distances is fast, it is possible that the
  metabolic rate of very small organisms is governed by this 
  process. It is well known that $L=D_0\tau^{1/2}$, where $D_0$ is the
 size-invariant diffusion coefficient.
 Since $\tau= (L/D_0)^{2}$,  we obtain from Eq. (\ref{eq1}) that
$M\propto L^{d+2}$. This relation furnishes how $L$ depends on $M$ and
we can use again Eq. (\ref{eq1}) to obtain that
\begin{eqnarray*}
L &\propto& M^\frac{1}{2+d}~,~~~~\tau\propto M^\frac{2}{2+d}~~,\\
B &\propto&  M^\frac{d}{2+d}~~.
\end{eqnarray*}
 In $d=3$, the metabolic exponent is $b=3/5$. 

 For larger organisms diffusion is inadequate. 
 Transport by convection is then utilized on large length scales. In mammals, for example,
 we find the cardiovascular system that transports blood to the capillaries, where the cells are
fed by diffusion. Since blood circulates in an ballistic regime,
 we consider that the BMR is basically driven by  
 {\sl ballistic transport}, namely $L=v_0\tau$, where the velocity $v_0$ is size-invariant.
  Now we must specify how the different metabolic steps are related.  
 We call BMR-1 the scenario of {\sl a single
 metabolic relevant length} $L_1\propto L=v_0\tau$ and {\sl a single time} $\tau$, both
  related to the ballistic transport.
  The other lengths, related to other metabolic steps, have evolved to meet
  it, namely $L_2\propto \ldots \propto L_d\propto L$. 
 Using that $V_d\propto L^d$ in Eq. (\ref{eq1}), we  write
\begin{eqnarray*}
L &\propto& \tau\propto M^\frac{1}{d+1}~~,\\
B &\propto&  M^\frac{d}{d+1}~~.
\end{eqnarray*}
 For $d=3$ we find the $3/4$-law, namely  $\tau\propto L\propto M^{1/4}$ and $B\propto M^{3/4}$.  
 These results are the same as those of WBE \cite{5} and Banavar et al. \cite{6,banavar2002}.
 
 In the BMR-2 scenario, we have  {\sl different relevant lengths and times} related to the metabolic 
 processes.
 However, due to the concept of symmorphosis \cite{15}, which states that all metabolic steps have
 co-evolved in order that no step is more limiting than another, we will end up with a single
 time $\tau$ and  $d-1$ rescaled lengths.
   In a ``cylindrical'' symmetry
  we have $L_1\propto L=v_0t_1$ (ballistic term) and $d-1$ lengths proportional to $R=D_0 t_2^{1/2}$
  (diffusion). Both $v_0$ and $D_0$ are size-invariant. From the symmorphosis principle 
    ($t_1=t_2=\tau$), it follows that $R=(D_0/v_0^{1/2})L^{1/2}$.
  The biological volume is $V_d\propto R^{d-1}L$.
 From Eq. (\ref{eq1}) we obtain that
\begin{eqnarray*}
L &\propto& \tau\propto M^\frac{2}{3+d}~~,~~~~R\propto M^\frac{1}{3+d}~~,\\
B &\propto&  M^\frac{1+d}{3+d}~~.
\end{eqnarray*}
 Then in $d=3$ the BMR-2 scenario yields the $2/3$ law,  without mention of the area/volume ratio. 
  We  obtained both  $3/4$ and $2/3$ laws from the same transport processes:
  convection and  diffusion. If convection is the dominant
 limiting process we have the $3/4$ law; 
   if the two processes are equivalent we obtain the $2/3$ law.

 The circulatory networks of endothermic animals are  dynamical ones which are adjusted
 according to the metabolic state. The transition from resting to maximum activity 
 can be described  very briefly as follows:
 (a) the heart increases its rate and output;
 (b) the mean arterial pressure and peripheral extramuscular resistance increase;
 (c) arterial blood volume increases due to constriction of the veins;
 (d) extramuscular flow remains essentially constant, somewhat reduced in some organs 
 but increased in others; and 
 (e) total flow and muscular flow increase, with all muscular capillaries activated. 
 The items (a), (c) and (e)
  suggest that we have a ``forced movement'' during the characteristic time  $\tau$.  
  This means that the typical constant velocity can be written as  $v=a_0\tau$, 
  where $a_0$ is a size-invariant acceleration.
  Consequently the MMR is driven by  an {\sl inertial movement accelerated during time $\tau$}, 
  implying that $L=v\tau=a_0\tau^2$. 
  
  If inertial transport is the only relevant  process (MMR-1 scenario), it follows that
  $L_1\propto L_2\ldots\propto L_d\propto L =a_0\tau^2$.
   Since $V_d\propto L^d$ and $\tau\propto L^{1/2}$, 
  we obtain from Eq. (\ref{eq1}) the metabolic relations: 
\begin{eqnarray*}
L &\propto& M^\frac{2}{2d+1}~~,~~~~\tau\propto M^\frac{1}{2d+1}~~,\\
B &\propto&  M^\frac{2d}{2d+1}~~.
\end{eqnarray*} 
 For $d=3$ we have that $L\propto M^{2/7}$, $\tau\propto M^{1/7}$  
 and $B\propto M^{6/7}$. 
  This results agree with the ones obtained trough a generalization of WBE ideas to
  MMR scenario \cite{12}.

 In the MMR-2 scenario, diffusion and inertial movement are equally relevant.
   We again choose a ``cylindrical'' symmetry so that 
 have that $L_1\propto L=a_0\tau^2$ while the remaining $d-1$ lengths are of order $D_0\tau^{1/2}$.
  It follows from Eq. (\ref{eq1}) that 
\begin{eqnarray*}
L &\propto& M^\frac{4}{d+5}~~,~~~~\tau\propto M^\frac{2}{d+5}~~,\\
B &\propto&  M^\frac{d+3}{d+5}~~, ~~~R\propto M^\frac{1}{d+5}~~.
\end{eqnarray*} 
 When $d=3$,  we obtain that $L\propto M^{1/2}$, $\tau\propto M^{1/4}$, $R\propto M^{1/8}$  
 and $B\propto M^{3/4}$.
 
\begin{table}
\caption{Allometric exponent $y$ describing the dependence of a variable $Y$ on body mass 
M ($Y\sim M^y$). Under parenthesis is the error in the last significative of the observed quantities.
 \label{tab1} }
\begin{center}
\begin{tabular}{|l|c|c|c|}
\hline
\multicolumn{1}{|c|} {\bf Variable} &  \multicolumn{3}{c|} {\bf Exponent}        \\
                           &     Predicted        &      Observed       & Ref. \\
\hline                           
                           &                      &                     &      \\
MMR                        &   $0.86$ (MMR-1)     &  $0.83(7)$          & \cite{8}    \\
                           &                      &  $0.88(2)$          &\cite{10}    \\
                           &                      &  $0.87(3)$          &\cite{11}    \\
                           &                      &  $0.85$             &\cite{20}   \\
                           &                      &  $0.87(5)$          &\cite{18}   \\ 
\hline                                                     
Capillary density            &   $-0.14$ (MMR-1)    &  $-0.14(7)$         &\cite{3,16} \\
\hline
Heart rate at MMR          &   $-0.14$ (MMR-1)    &  $-0.17(2)$         &\cite{21}   \\
                           &                      &  $-0.16(2)$         &\cite{22}   \\
                           &                      &  $-0.15$            &\cite{23}   \\
\hline
BMR                        &      $0.75$  (BMR-1) & $0.74(2)$     &\cite{8} \\    
                           &      $0.66$  (BMR-2) & $0.67$        &\cite{7}   \\ 
                           &                      & $0.69(1)$     &\cite{24} \\                                                                           
\hline
Heart rate at BMR          &     $-0.25$  (BMR-1) &  $-0.25(2)$   &\cite{1,8}   \\
                           &     $-0.33$ (BMR-2)  &  $-0.27$      &\cite{3,24}    \\
\hline
Aorta radius               &      $0.36$ (MMR-1)  &  $0.36$       &\cite{3,24}\\
                           &      $0.38$ (BMR-1)  &               &   \\
                           &      $0.33$ (BMR-2)  &               &   \\                          
\hline
Aorta length               &      $0.29$ (MMR-1)  &  $0.32$       &\cite{1,3}   \\ 
                           &      $0.25$ (BMR-1)  &  $0.31$       &\cite{24}   \\           
                           &      $0.33$ (BMR-2)  &               &  \\                           
\hline
\end{tabular}
\end{center}
\end{table}

 The transportation network can be characterized by  ``aorta'' and ``capillaries''.
 Note that the aorta $L_a$ and capillary $l_c$  lengths are both proportional to $L$.
  Since the nutrient fluid is conserved, the volume rate of flow is given by
\[  
{\dot Q}=\pi R_a^2v_a=N_c\pi r_c^2v_c~~,
\]
 where $R_a$ and $v_a$ are the aorta radius and fluid velocity and $N_c$, $r_c$ and $v_c$ are
 capillary number, radius and fluid velocity, respectively. It is natural to write that
 ${\dot Q}\propto B$. In the basal regime, $v_a$ and $v_c$ are size-invariant. 
 Then we obtain that $R_a\propto B^{1/2}$ and
 $N_cr_c^2\propto B$. Making the {\sl extra assumption} that $r_c$ is invariant, it follows
 that the capillary density $\rho_c=N_c/M$ is $\rho_c\propto B/M\propto 1/\tau$. For $d=3$
 we have the following results: (a) BMR-1 - $R_a\propto M^{3/8}$  and $\rho_c\propto M^{-1/4}$;
  (b) BMR-2 - $R_a\propto M^{1/3}$  and $\rho_c\propto M^{-1/3}$.
  In the maximum regime, we have that $v_a=a_0\tau$, implying that now $v_a$ depends 
  on the mass. Note that this suggests new empirical studies. Then, we obtain 
  that $R_a^2\propto B/\tau$. Now $v_c$ is not necessarily invariant, as in the basal case. 
  Since $\rho_cr_c^2v_c\propto 1/\tau$ we obtain that $\rho_c\propto 1/\tau$ if we make 
  the {\sl extra assumptions} that both $v_c$ and $r_c$ are independent of body mass. 
  The results are: (c) MMR-1 - $R_a\propto M^{5/14}$ and $\rho_c\propto M^{-1/7}$;
 (d) MMR-2 - $R_a\propto M^{1/4}$ and $\rho_c\propto M^{-1/4}$.

 Let us compare our predictions for $d=3$ with experimental data (see Table \ref{tab1}).
  The values predicted in the  MMR-2 context are far from the experimental ones.  
 On the other hand, the MMR-1 scenario describes very well the MMR data. 
 The exponent $b=6/7\approx 0.86$,
 larger than the basal value, is in very good agreement with data. 
 Muscular capillary density of mammals is linked to MMR, instead of BMR,  
  because only during exercise  are all the muscular capillaries perfused. 
  The capillary density scales as $\rho_c=N_c/M\propto M^{-1/7}$, in
 good agreement with the average experimental value for
 various regions of muscle \cite{3,16}.
 Since $\tau\propto M^{1/7}$, frequencies must scale as $F\propto \tau^{-1}\propto M^{f}$ 
 with $f=-1/7\approx -0.14$. This value, smaller than the basal one, is also in good
 agreement with data for heart and respiration rates in strenuous exercise. 
 The results for the capillary  invariant radius $r_c$  and $l_c\propto M^{2/7}$,
  agree roughly with the theoretical-empirical estimation of Dawson \cite{17}. 
  ($l_c\propto M^{0.21}$ and $r_c\propto M^{0.08}$). 
  
  Since the length of aorta cannot change from basal to maximum metabolic regimes,
  it should scale as the prediction of MMR scaling. The predicted exponent $0.29$ agrees well
  with data. The aorta radius could  in principle follow the two scalings because of
  the elasticity  of the
  aorta and  the dynamical body adaptations of mammals in the transition BMR - MMR.
  The experimental value $0.36$ has however a better agreement with the MMR-1 value $5/14$.  

 Consider now the predictions of BMR scaling. Recently the empirical values of the
 BMR exponent of mammals were reanalyzed using diverse  procedures \cite{7,8,18} that furnished
 values in the interval between $b_2=2/3$ and $b_1=3/4$, which are the predicted values
  of BMR-2 and BMR-1, respectively. Heart and respiration
 rates are close to the BMR-1 value $-1/4$ and other biological variables \cite{5} have
 values close to multiples of $1/4$. On the other hand, empirical data near multiples of
 $1/3$ are also reported \cite{9}. Therefore, the two scenarios are possible.
 This last possibility explains why $b$ is greater in large versus small mammals data:
  diffusion and ballistic transports can be equally important in small organisms (BMR-2) 
  but not in large ones, where ballistic transport is crucial (BMR-1).
   Finally, let us emphasize that we make a verifiable empirical prediction: the aorta blood
   velocity ($v_a$), which is scaling-invariant in BMR conditions, grows with mass in the
   exercise-induced MMR condition ($v_a\propto \tau$). The related exponent, which is predict
   to have the value $-0.14$ in the MMR-1 scenario, was never measured. Its empirical determination
   can be an experimental test of the importance of the transportation processes for the allometric
   scaling of metabolism.

\noindent{\bf Acknowledgements - }
  We are grateful to R. Dickman and S. O. Kamphorst for 
 careful readings of the manuscript. JKLS thanks A. Maritan for
 useful  discussions by e-mail.
 We thank CNPq, CAPES and FAPEMIG, Brazil, for financial support. 
 LAB was also supported in part by FAPESP, Brazil.

\end{document}